\documentclass{iaus}
\usepackage{graphicx}
\usepackage{natbib}

\def\solmas{$\mathrm{M_\odot}$~}

\def\simless{\mathbin{\lower 3pt\hbox
   {$\rlap{\raise 5pt\hbox{$\char'074$}}\mathchar"7218$}}}
\def\simgreat{\mathbin{\lower 3pt\hbox
   {$\rlap{\raise 5pt\hbox{$\char'076$}}\mathchar"7218$}}}
\newcommand{\zfrag}{${\rm Z} = 10^{-5} \: {\rm Z_{\odot}} $}
\newcommand{\zsix}{${\rm Z} = 10^{-6} \: {\rm Z_{\odot}} $}

\title[Formation of Stellar Clusters] 
{Formation of Stellar Clusters and the Importance of Thermodynamics for Fragmentation}

\author[Klessen, Clark, Glover]   
{Ralf S.\ Klessen$^1$, Paul C.~Clark$^1$, Simon C.~O.~Glover$^2$}%
 
\affiliation{$^1$Zentrum f{\"u}r Astronomie der Universit{\"a}t Heidelberg, Institut f{\"u}r Theoretische Astrophysik, Albert-Ueberle-Str.\ 2, 69120 Heidelberg, Germany \\[\affilskip]
$^2$Astrophysikalisches Institut Potsdam, An der Sternwarte 16, 14482 Potsdam,
Germany  
}

\pubyear{2004}
\volume{xxx}  
\pagerange{119--126}
\date{?? and in revised form ??}
\jname{Proceedings Title IAU Symposium}
\editors{A.C. Editor, B.D. Editor \& C.E. Editor, eds.}
\begin{document}

\maketitle

\begin{abstract}
We discuss results from numerical simulations of star cluster formation in the turbulent interstellar medium (ISM). The thermodynamic behavior of the star-forming gas plays a crucial role in fragmentation and determines the stellar mass function as well as the dynamic properties of the nascent stellar cluster. This holds for star formation in molecular clouds in the solar neighborhood as well as for  the formation of the very first stars in the early universe.  The thermodynamic state of the ISM is a result of the balance between 
heating and cooling processes, which in turn are determined 
by atomic and molecular physics and by chemical 
abundances. Features in the effective equation of state of the gas, such as a transition from a cooling to a heating regime, define a characteristic mass scale for fragmentation and so set the peak of the initial mass function of stars (IMF).  As it is  based on  fundamental physical quantities and constants,  this is an attractive approach to explain the apparent universality of the IMF in the solar neighborhood as well as the transition from purely primordial high-mass star formation to the more normal low-mass mode observed today.
\end{abstract}

\firstsection 

\section{Introduction} 

Identifying the physical processes that determine the masses
of stars and their statistical distribution, the initial mass function
(IMF), is a fundamental problem in star-formation research. It is
central to much of modern astrophysics, with implications ranging from
cosmic re-ionisation and the formation of the first galaxies, over the
evolution and structure of our own Milky Way, down to the build-up of
planets and planetary systems.

Near the Sun the number density of stars as a function
of mass has a peak at a characteristic stellar mass of a few tenths of
a solar mass, below which it declines steeply, and for masses above
one solar mass it follows a power-law with an exponent $dN/d{\log}m
\propto m^{-1.3}$. Within a radius of several kpc 
this distribution shows surprisingly little variation (Salpeter 1955;
Scalo 1998; Kroupa 2001; Kroupa 2002; Chabrier 2003).  This has
prompted the suggestion that the distribution of stellar masses at
birth is a truly universal function, which often is referred to as the
Salpeter IMF, although note that the original Salpeter (1955) estimate
was a pure power-law fit without characteristic mass scale. 

The initial conditions in star forming
regions can vary considerably, even in the solar vicinity. If the IMF were to depend on the initial
conditions, there would be no reason for it to be universal.
Therefore a derivation of the characteristic stellar mass that is
based on fundamental atomic and molecular physics would be
highly desirable. In this proceedings contribution we argue that indeed the thermodynamic properties of the star-forming cloud material determine the characteristic mass scale for fragmentation and subsequent stellar birth.   The thermodynamic state of interstellar gas is a result of
the balance between heating and cooling processes, which in turn are
determined by fundamental atomic and molecular physics and by chemical abundances. The derivation
of a characteristic stellar mass can thus be based on quantities and constants
that depend solely on the chemical abundances in a molecular cloud. It also explains why deviations from the ``standard'' mode of star formation are likely to occur under extreme environmental conditions such as those occurring in the early universe or in circum-nuclear starburst regions.

\section{Gravoturbulent Star Cluster Formation}

Stars and star clusters form through the interplay between
self-gravity on the one hand and turbulence, magnetic fields, and
thermal pressure on the other
(for recent reviews see Larson 2003; Mac~Low \& Klessen 2004;
Ballesteros-Paredes et al.\ 2006). 
Supersonic turbulence, even if it is strong enough to counterbalance
gravity on global scales, will usually {\em provoke} local collapse.
Turbulence establishes a complex network of interacting shocks, where
converging shock fronts generate clumps of high density. These density
enhancements can be large enough for the fluctuations to become
gravitationally unstable and collapse, which can occur when the local
Jeans length becomes smaller than the size of the fluctuation.
However, the fluctuations in turbulent velocity fields are highly
transient.  The random flow that creates local density enhancements
can disperse them again.  For local collapse to actually result in the
formation of stars, Jeans-unstable shock-generated density
fluctuations must collapse to sufficiently high densities on time
scales shorter than the typical time interval between two successive
shock passages.  Only then are they able to `decouple' from the
ambient flow and survive subsequent shock interactions.  The shorter
the time between shock passages, the less likely these fluctuations
are to survive. Hence, the timescale and efficiency of protostellar
core formation depend strongly on the wavelength and strength of the
driving source as well as on the dynamic  response of the gas, as defined by the equation of state, i.e. the balance between heating and cooling processes.

The velocity field of long-wavelength turbulence is found to be
dominated by large-scale shocks which are very efficient in sweeping
up molecular cloud material, thus creating massive coherent
structures. When a coherent region reaches the critical density for
gravitational collapse, its mass typically exceeds the local Jeans
limit by far.  Inside the shock compressed region, the velocity
dispersion is much smaller than in the ambient turbulent flow and the
situation is similar to localized turbulent decay.  These are the conditions for the formation of star clusters.  The efficiency of
turbulent fragmentation is reduced if the driving wavelength
decreases. When energy is inserted mainly on small spatial scales, the
network of interacting shocks is very tightly knit, and protostellar
cores form independently of each other at random locations throughout
the cloud and at random times.  Individual shock-generated clumps have
lower mass and the time interval between two shock passages through
the same point in space is small.  Collapsing cores are easily
destroyed again and the resulting mass spectrum shows deviations from the observed IMF. All this points toward interstellar gas clouds being driven on large scales. 

Altogether, stellar birth is intimately
linked to the dynamic behavior of the parental gas cloud, which
governs when and where star formation sets in.
The chemical and thermodynamic properties of interstellar clouds play
a key role in this process. In particular, the value of the polytropic
exponent $\gamma$, when adopting an EOS of the
form $P\propto\rho^\gamma$, strongly influences the compressibility of
density condensations as well as the temperature of the gas. The EOS
thus determines the amount of clump fragmentation, and so directly
relates to the IMF (V\'azquez-Semadeni et al.\ 1996) with values of
$\gamma$ larger than unity leading to little fragmentation and high
mass cores (Li, Klessen, \& Mac~Low 2003; Jappsen et al.\ 2005). The
stiffness of the EOS in turn depends strongly on the ambient
metallicity, density and infrared background radiation field produced
by warm dust grains. The EOS thus varies considerably in different
galactic environments (see Spaans \& Silk 2000, 2005 for a detailed
account).

\section{Formation of Stellar  Clusters in the Solar Neighborhood}

Early studies of the balance between heating and cooling processes in
collapsing clouds predicted temperatures of the order of $10\,$K to $20\,$K,
tending to be lower at the higher densities
\citep[e.g.,][]{HAY65,HAY66,LAR69,LAR73}. In their dynamical collapse calculations,
these and other authors approximated this somewhat varying temperature by
a simple constant value, usually taken to be $10\,$K.  Nearly all subsequent
studies of cloud collapse and fragmentation have used a similar isothermal
approximation.  However, this approximation is actually only a somewhat
crude one, valid only to a factor of 2, since the temperature is predicted
to vary by this much above and below the usually assumed constant value
of $10\,$K.  Given the strong sensitivity of the results of fragmentation
simulations like those of \citet{LI03} to the assumed
equation of state of the gas, temperature variations of this magnitude may
be important for quantitative predictions of stellar masses and the IMF.

As can be seen in Fig.~2 of \citet{LAR85}, observational and theoretical
studies of the thermal properties of collapsing clouds both indicate that
at densities below about $10^{-18}\,\mathrm{g\,cm}^{-3}$, roughly corresponding
to a number density of $n = 2.5\times 10^5\,\mathrm{cm}^{-3}$, the temperature generally
decreases with increasing density.  In this low-density regime, clouds are
externally heated by cosmic rays or photoelectric heating, and they are
cooled mainly by the collisional excitation of low-lying levels of C$^+$ ions and
O atoms; the strong dependence of the cooling rate on density then yields
an equilibrium temperature that decreases with increasing density.  The
work of \citet{KOY00}, which assumes that photoelectric heating
dominates, rather than cosmic ray heating as had been assumed in earlier
work, predicts a very similar trend of decreasing temperature with
increasing density at low densities.  
The three-dimensional  
magnetohydrodynamic simulations of \citet{GM07} also produce  
a similar result, although in this case the point-to-point scatter is larger.
The resulting temperature-density
relation can be approximated by a power law with an exponent of about
$-0.275$, which corresponds to a polytropic equation of state with
$\gamma = 0.725$.  The observational results of \citet{MYE78} shown in Fig.~2
of \citet{LAR85} suggest temperatures rising again toward the high end of
this low-density regime, but those measurements refer mainly to relatively
massive and warm cloud cores and not to the small, dense, cold cores in
which low-mass stars form.  As reviewed by \citet{EVA99}, the temperatures
of these cores are typically only about $8.5\,$K at a density of
$10^{-19}\,\mathrm{g\,cm}^{-3}$, consistent with a continuation of the decreasing trend noted
above and with the continuing validity of a polytropic approximation with
$\gamma \approx 0.725$ up to a density of at least $10^{-19}\,\mathrm{g\,cm}^{-3}$.

At higher densities, atomic line cooling becomes less effective as the cooling
rates start to reach their local thermodynamic equilibrium (LTE) limits and as
the line opacities grow larger. Consequently,
at densities above $10^{-18}\,\mathrm{g\,cm}^{-3}$ the gas becomes
thermally coupled to the dust grains, which then control the temperature by
their far-infrared thermal emission.  In this high-density regime, dominated
thermally by the dust, there are few direct temperature measurements because
the molecules normally observed freeze out onto the dust grains, but most of
the available theoretical predictions are in good agreement concerning the
expected thermal behavior of the gas \citep{LAR73, LOW76, MAS00,LAR05}. 
The balance between compressional heating and
thermal cooling by dust results in a temperature that increases slowly with
increasing density, and the resulting temperature-density relation can be
approximated by a power law with an exponent of about $0.075$, which
corresponds to $\gamma = 1.075$.  Between the low-density and the high-density
regimes, the temperature is predicted to reach a minimum of $5\,$K at a density
of about $2 \times 10^{-18}\,\mathrm{g\,cm}^{-3}$, at which point the Jeans
mass is about $0.3\,M_{\odot}$. The actual minimum temperature reached is somewhat uncertain
because observations have not yet confirmed the predicted very low values,
but such cold gas would be very difficult to observe; various efforts to
model the observations have suggested central temperatures between $6\,$K and
$10\,$K for the densest observed prestellar cores, whose peak densities may
approach $10^{-17}\,\mathrm{g\,cm}^{-3}$ \citep[e.g.][]{ZUC01, EVA01, TAF04}.  A power-law approximation to the equation of state
with $\gamma \approx 1.075$ is expected to remain valid up to a density of about
$10^{-13}\,\mathrm{g\,cm}^{-3}$, above which increasing opacity to the thermal emission
from the dust causes the temperature to begin rising much more rapidly,
resulting in an ``opacity limit'' on fragmentation that is somewhat below
$0.01\,M_{\odot}$ \citep{LOW76, MAS00}.

Adopting the piecewise polytropic equation of state outlined above, Jappsen et al. (2005) showed that  the changing $\gamma$ from a value below unity to one somewhat above unity at a 
critical density $n_{\rm c}$ influences the number of protostellar objects. 
If the critical density increases then more protostellar objects 
form but the mean mass decreases.
Consequently, the 
peak of the resulting mass spectrum moves to lower masses with increasing 
critical density. This spectrum not only shows a pronounced 
peak but also a powerlaw tail towards higher masses. Its behavior is thus similar to the observed IMF.

A simple scaling argument 
based on the Jeans mass $M_{\rm J}$ at the critical density $n_{\rm c}$ leads to $M_{\rm ch} \propto n_{\rm c}^{-0.95}$.  If there is a close relation between the average 
Jeans mass and the characteristic mass of a fragment, a similar 
relation should hold for the expected peak of the mass spectrum. The  simulations by Jappsen et al. (2005) qualitatively support this hypothesis, 
but find a weaker density dependency $M_{\rm ch} \propto n_{\rm c}^{-0.5\pm 0.1}$

\section{Formation of Stellar  Clusters in the Early Universe}

The formation of the first and second generations of stars  in the early universe has far-reaching consequences for cosmic reionization and galaxy formation \cite[][]{lb01,bl04,glover05}. The physical processes that govern stellar birth in a metal-free or metal-poor environment, however, are still very poorly understood. Numerical simulations of the thermal and dynamical evolution of gas in primordial protogalactic halos indicate that the metal-free first stars, the so called Population III, are expected to be very massive, with masses 
anywhere in the range 20--$2000\,{M_{\odot}}$ \cite[][]{abn02,bcl02,yoha06,oshn07},
much larger than the characteristic mass scale found in the local IMF.

This means that at some stage of cosmic evolution there must have been a transition from primordial, high-mass star formation to the ``normal'' mode of star formation that dominates today. 
The discovery of extremely metal-poor subgiant stars in the Galactic halo with masses below one solar mass \cite[][]{christlieb02,bc05} indicates that this transition occurs at abundances considerably smaller  than the solar value. At the extreme end, these stars have iron abundances less than $10^{-5}\,$Z$_{\odot}$,  and carbon or oxygen abundances that are still $ \simless \,10^{-3}$ the solar value. These stars are thus strongly  iron deficient, which  could be due to unusual abundance patterns produced by enrichment from pair-instability supernovae  \citep{hw02} from Population III stars or due to mass transfer from a close binary companion \citep{ry05,kom07}. There are hints for an increasing binary fraction with decreasing metallicity for these stars \cite[][]{luc05}.

If metal enrichment is the key to the formation of low-mass stars, then logically
there  must be some critical metallicity ${\rm Z_{\rm crit}}$ at which the formation
of low mass stars first becomes possible. However, the value of  ${\rm Z_{\rm crit}}$
is a matter of ongoing debate. Some models suggest that low mass star formation
becomes possible only once atomic fine-structure line cooling from carbon and oxygen
becomes effective \citep{bfcl01,bl03,san06,fjb07}, setting a value for ${\rm Z_{\rm
crit}}$ at around $10^{-3.5} \: {\rm Z_{\odot}}$. Another possibility  is that low mass star formation is a result of
dust-induced fragmentation occurring at high densities, and thus at a very late stage
in the protostellar collapse \citep{sch02,om05,sch06,to06}. In this model,
$10^{-6}~\simless~{\rm Z_{\rm crit}}~\simless~10^{-4}~{\rm Z_{\odot}}$, where much of
the uncertainty in the predicted value results from uncertainties in the dust
composition and the degree of gas-phase depletion \citep{sch02,sch06}.
%

\begin{figure*}[t]
\begin{center}
\includegraphics[width=3.1in,height=1.75in]{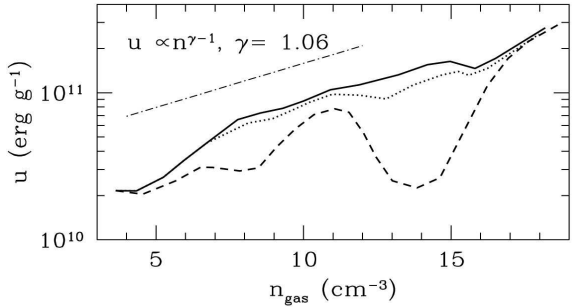}
\end{center}
\caption{\label{fig:EOS} Three equations of state (EOSs) from Omukai et~al. (2005) for different metallicities used by \cite{cgk07}.  The primordial case (solid line), \zsix~(dotted line), and
\zfrag~(dashed line), are shown alongside an example of a polytropic EOS with an
effective $\gamma = 1.06$. }
\end{figure*}

\begin{figure}
\begin{center}
\unitlength1cm
\begin{picture}(11.6,17.4)
\put(0.0, 11.6){\includegraphics[width=5.8cm,height=5.8cm]
			{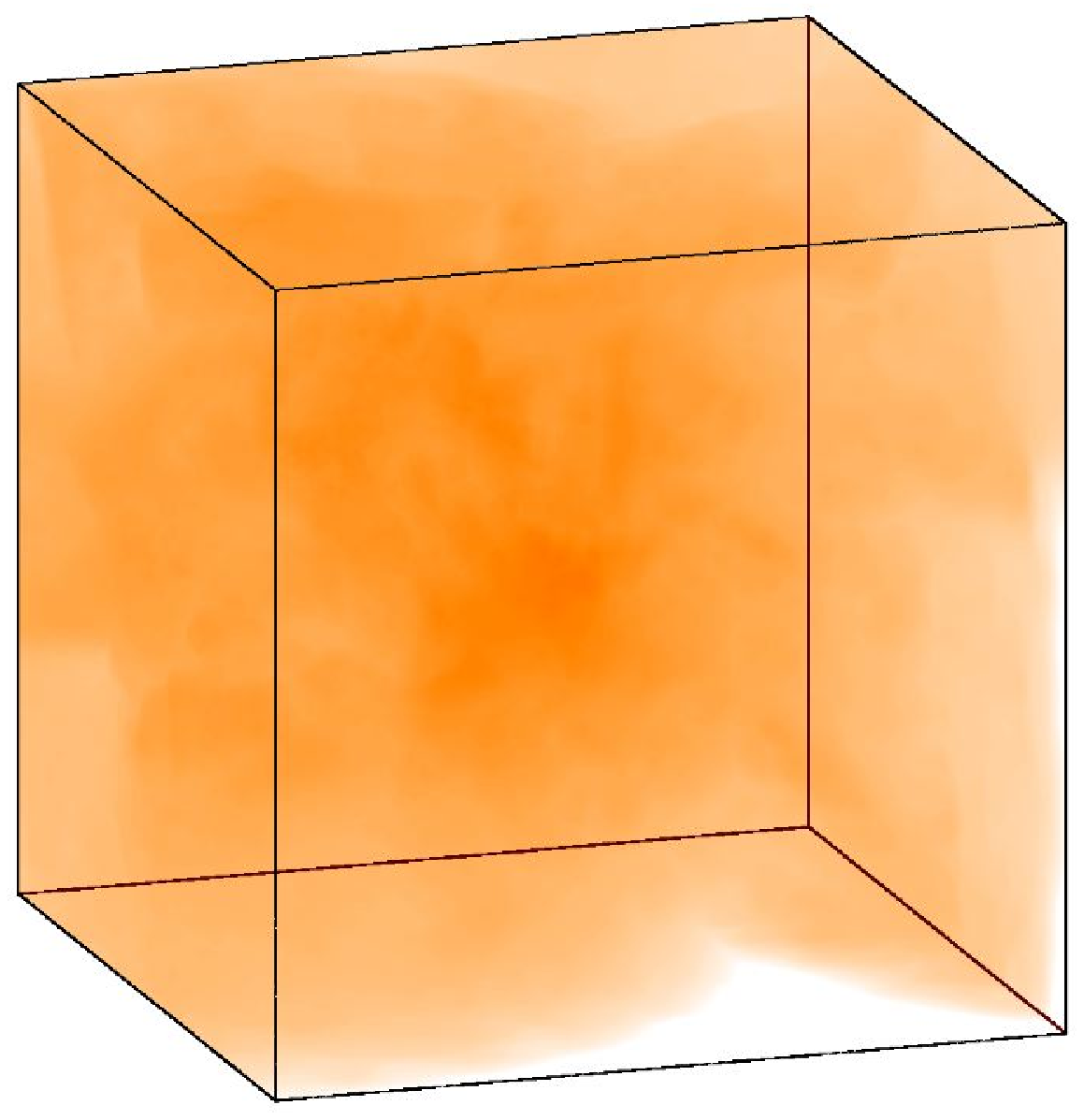}}
\put(5.8, 11.6){\includegraphics[width=5.8cm,height=5.8cm]
			{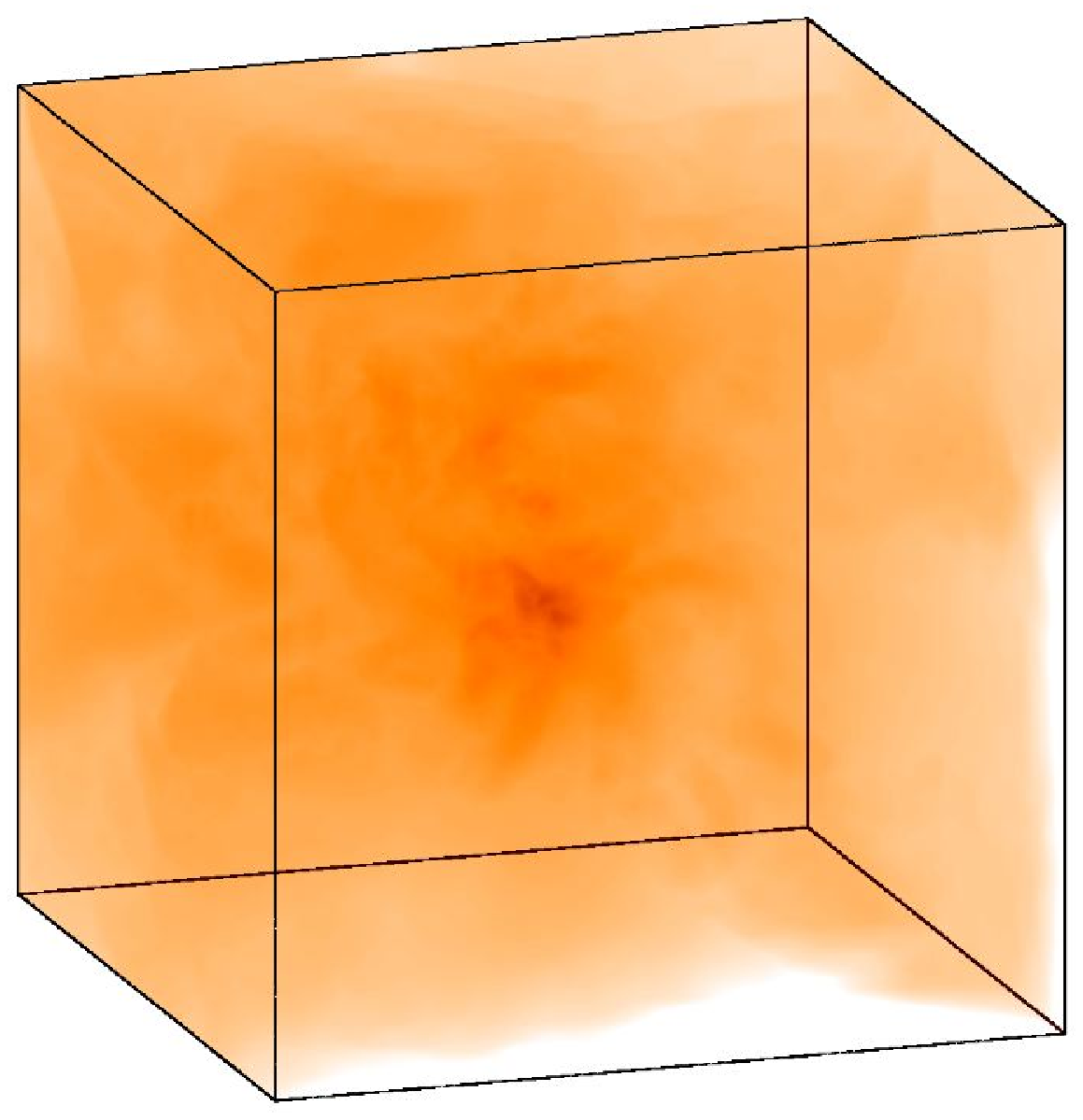}}
\put(0.0,  5.8){\includegraphics[width=5.8cm,height=5.8cm]
			{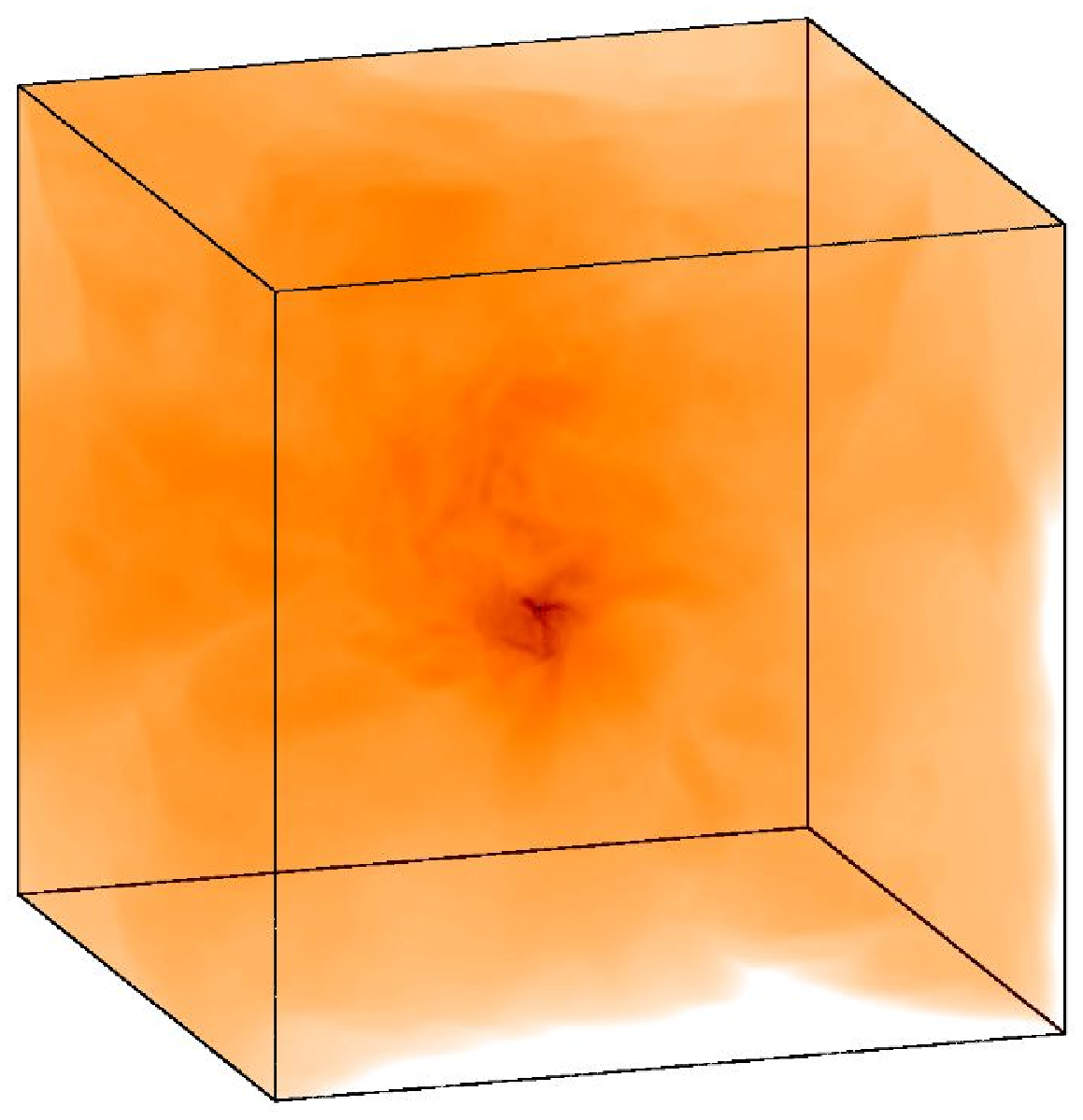}}
\put(5.8,  5.8){ \includegraphics[width=5.8cm,height=5.8cm]
			{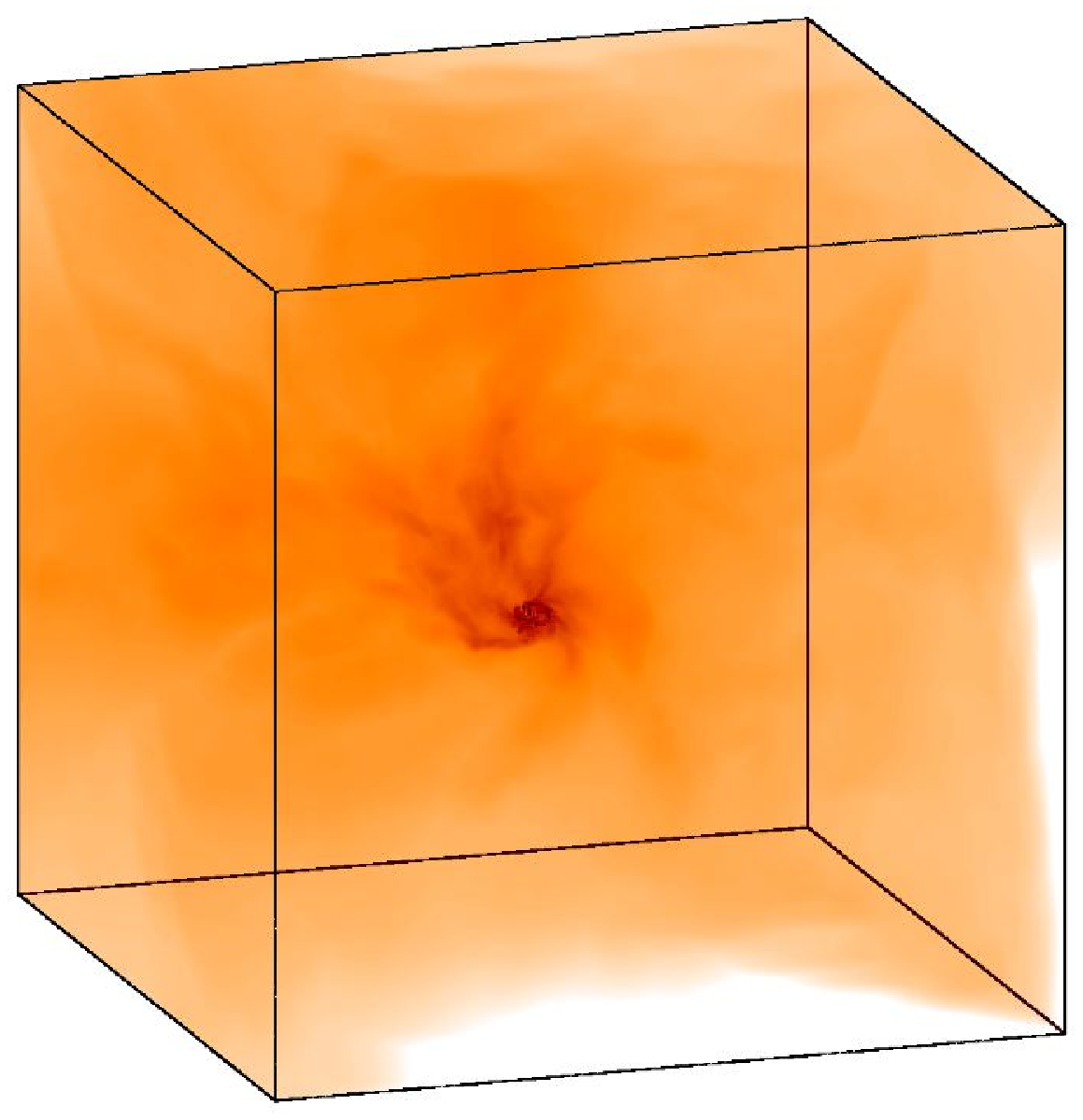}}
\put(0.0,  0.0){\includegraphics[width=5.8cm,height=5.8cm]
			{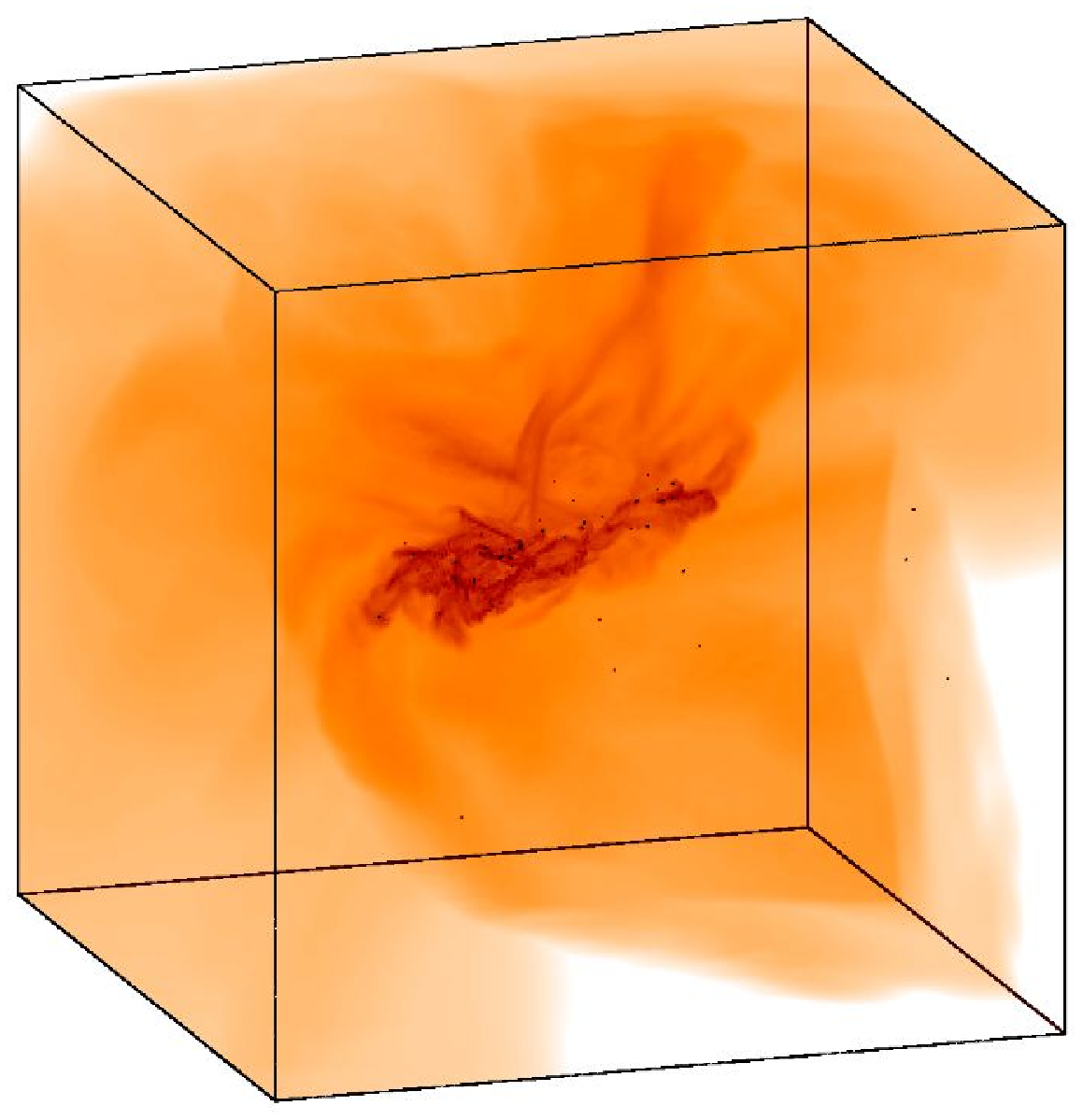}}
\put(5.8,  0.0){\includegraphics[width=5.8cm,height=5.8cm]
			{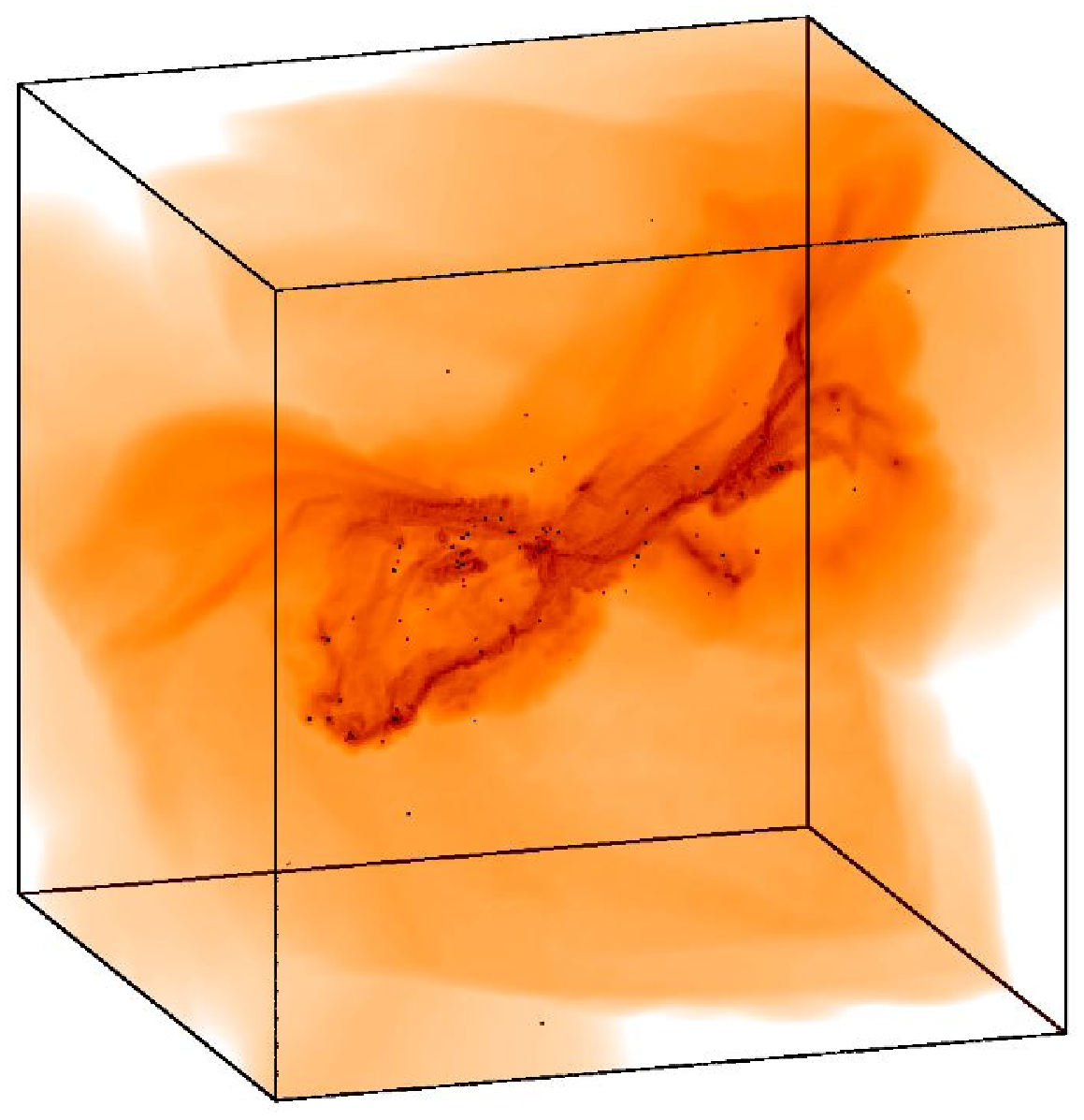}}
\put(1.9, 11.9){\Large {$t = t_{\rm SF} - 67$ yr}}
\put(7.7, 11.9){\Large {$t = t_{\rm SF} - 20$ yr}}
\put(1.9, 6.1){\Large {$t = t_{\rm SF}$ }}
\put(7.7, 6.1){\Large {$t = t_{\rm SF} + 53$ yr}}
\put(1.9, 0.3){\Large {$t = t_{\rm SF} +233$ yr}}
\put(7.7, 0.3){\Large {$t = t_{\rm SF} +420$ yr}}
\end{picture}
\end{center}
\caption{\label{fig:sequence} Time evolution of the density distribution in the innermost 400 AU of the
protogalactic halo shortly before and shortly after the formation of the first protostar at
$t_{\rm SF}$. Only gas at densities above $10^{10}\,$cm$^{-3}$ are plotted.  The dynamical
timescale at a density $n = 10^{13}\,$cm$^{-3}$ is of the order of only 10 years.  Dark dots
indicate the location of protostars as identified by sink particles forming at $n \ge
10^{17}\,$cm$^{-3}$. Note that without usage of sink particles to identify collapsed protostellar cores one  would not have been able
to follow the build-up of the protostellar cluster beyond the formation of the first object.
There are 177 protostars when we stop the calculation at $t = t_{\rm SF} + 420\,$yr. They
occupy a region roughly a hundredth of the size of the initial cloud. With
$18.7\,$M$_{\odot}$  accreted at this stage, the stellar density is $2.25 \times
10^{9}\,$M$_{\odot}\,$pc$^{-3}$. Data are from \cite{cgk07}.}

\end{figure}

\begin{figure}
\begin{center}
\unitlength1cm
\begin{picture}(18,5.5)
\put(1.0, 0.50){\includegraphics[width=1.2in,height=1.2in]{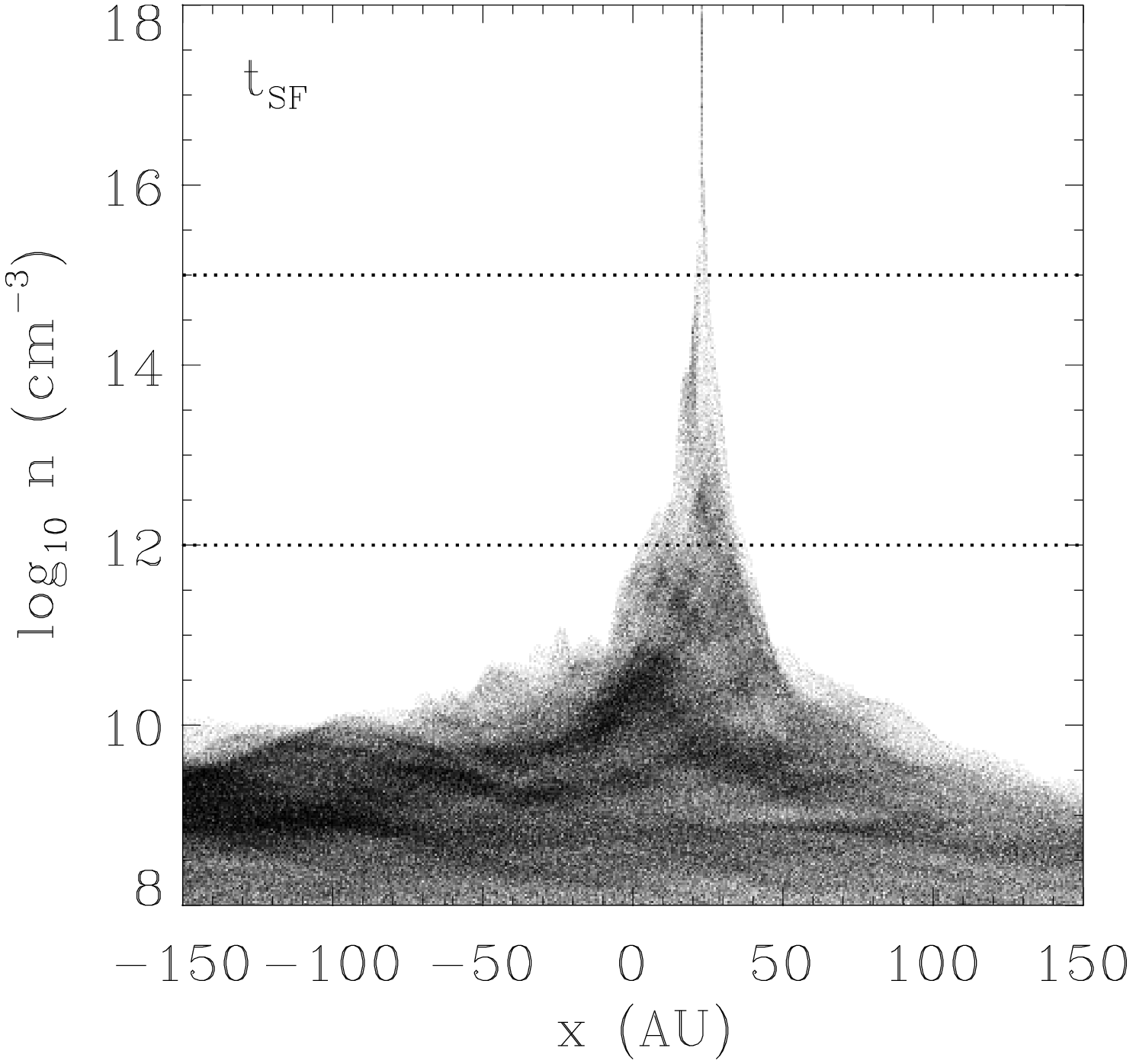}}
\put(5.5, 0.50){\includegraphics[width=1.2in,height=1.2in]{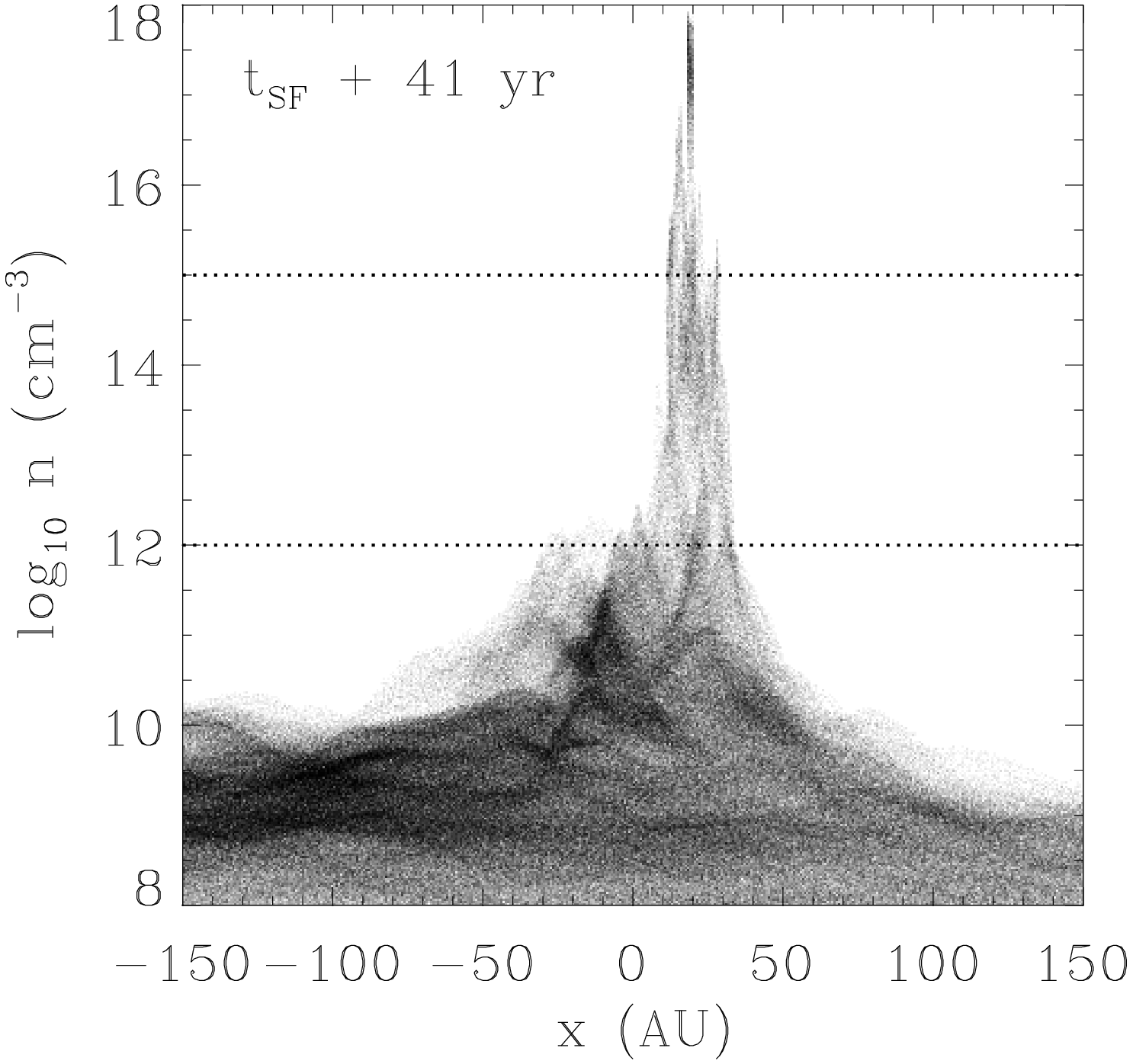}}
\put(10.0, 0.50){\includegraphics[width=1.2in,height=1.2in]{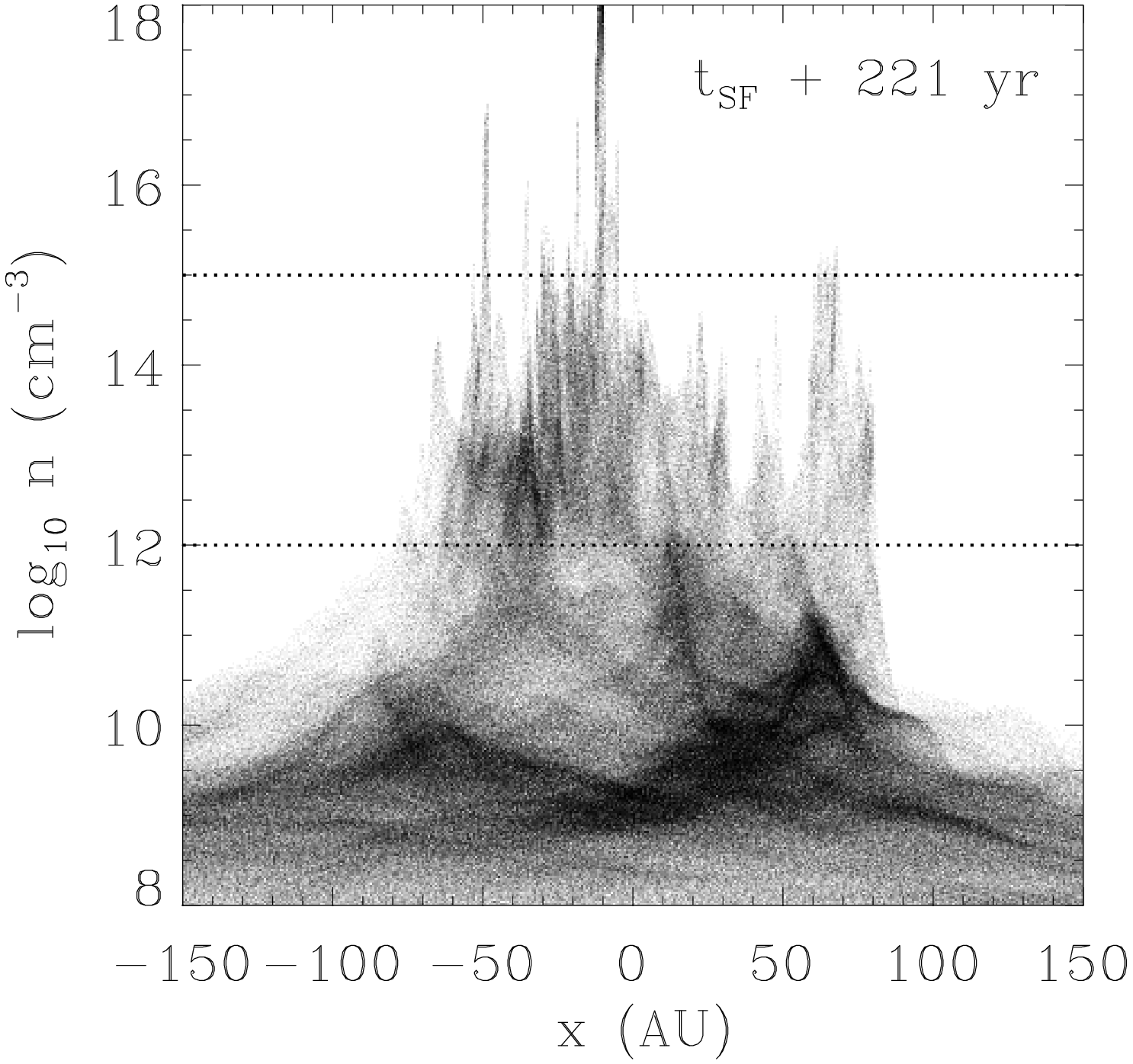}}
\end{picture}
\end{center}
\caption{ \label{fig:density} To illustrate the onset of the fragmentation process in the 
\zfrag~simulation, the graphs show the densities of the particles, plotted as a
function of their x-position. Note that for each plot, the particle data has been centered on
the region of interest.  Results are plotted for three different output times, ranging from
the time that the first star forms ($t_{\rm sf}$) to 221 years afterwards. The densities 
lying between the two horizontal dashed lines denote the range over which
dust cooling lowers the gas temperature. The figure is from \cite{cgk07}.}
\end{figure}

\cite{cgk07} modeled star formation in the central regions of low-mass halos at high redshift adopting an EOS similar to Omukai et al.\ (2005). They focused on a high-density regime with $10^5\,$cm$^{-3} \le n \le 10^{17}\,$cm$^{-3}$. They find that enrichment of the gas to a metallicity of only ${\rm Z} = 10^{-5} \:{\rm
Z_{\odot}}$ dramatically enhances fragmentation. A typical time evolution is 
 illustrated in Figure
\ref{fig:sequence}. It shows several stages in the collapse process, spanning a time interval from
shortly before the formation of the first protostar (as identified by the formation of
a sink particle in the simulation) to 420 years afterwards. 
During the initial contraction, the cloud builds up a central core with a density of
about $n = 10^{10}\,$cm$^{-3}$. This core is supported by a combination of thermal
pressure and rotation. Eventually, the core reaches high enough densities to go into
free-fall collapse, and forms a single protostar. As more high angular momentum
material falls to the center, the core evolves into a disk-like structure with density
inhomogeneities caused by low levels of turbulence.  As it grows in mass, its density
increases. When dust-induced cooling sets in, it fragments heavily into a tightly
packed protostellar cluster within only a few hundred years. One can see this behavior
in particle density-position plots in Figure \ref{fig:density}. The  simulation is stopped
420 years after the formation of the first stellar object (sink particle). At this
point, the core has formed 177 stars. The evolution in the low-resolution simulation is
very similar. The time between the formation of the first and second protostars is
roughly 23 years, which is two orders of magnitude higher than the free-fall time at
the density where the sinks are formed. Note that without the inclusion of sink particles, one 
would only have been able to capture the formation of the first collapsing object which
forms the first protostar: the formation of the accompanying cluster would have
been missed entirely.

\begin{figure}
\centerline{
	\includegraphics[width=1.65in,height=1.65in]{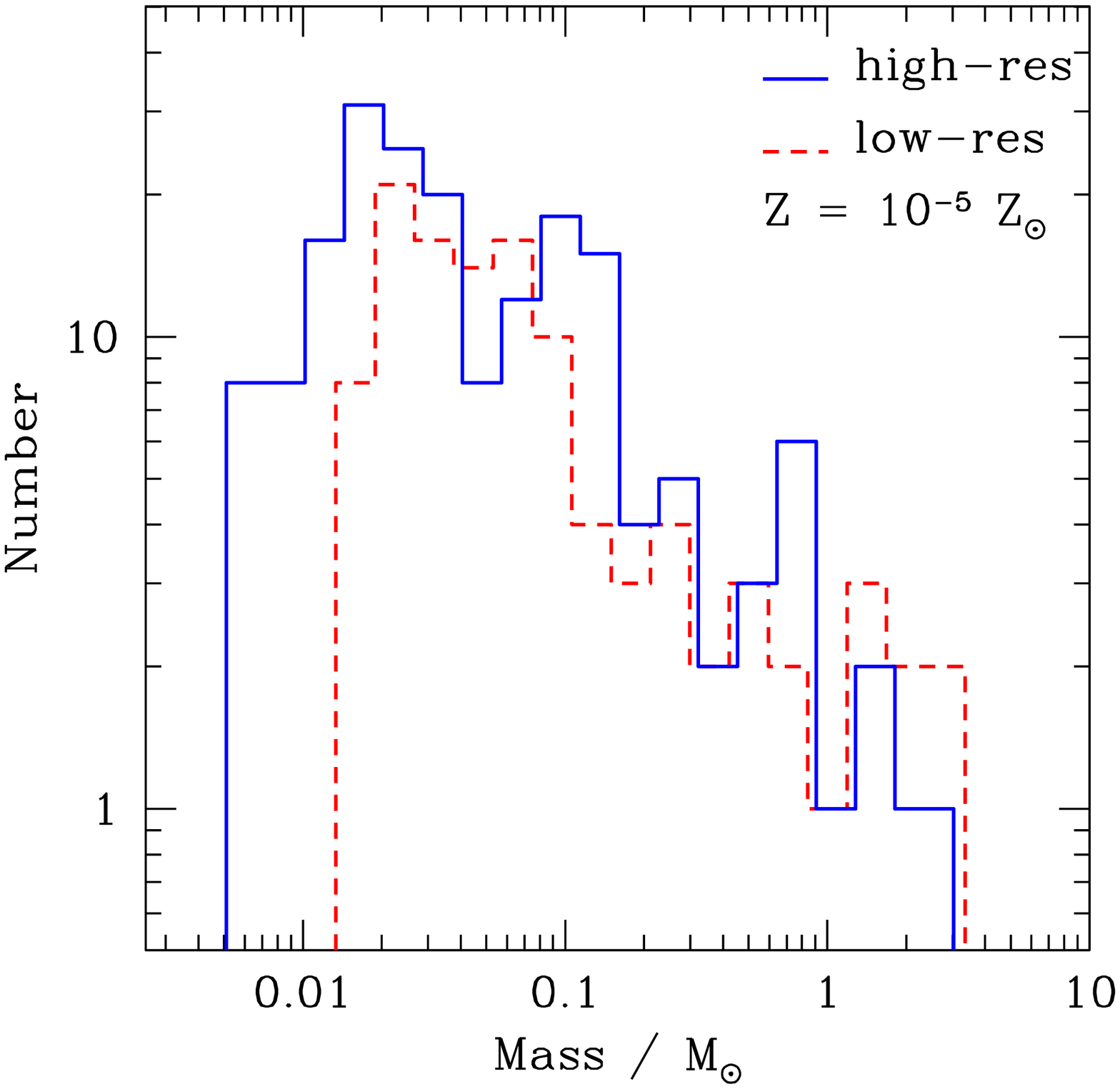}
	\includegraphics[width=1.65in,height=1.65in]{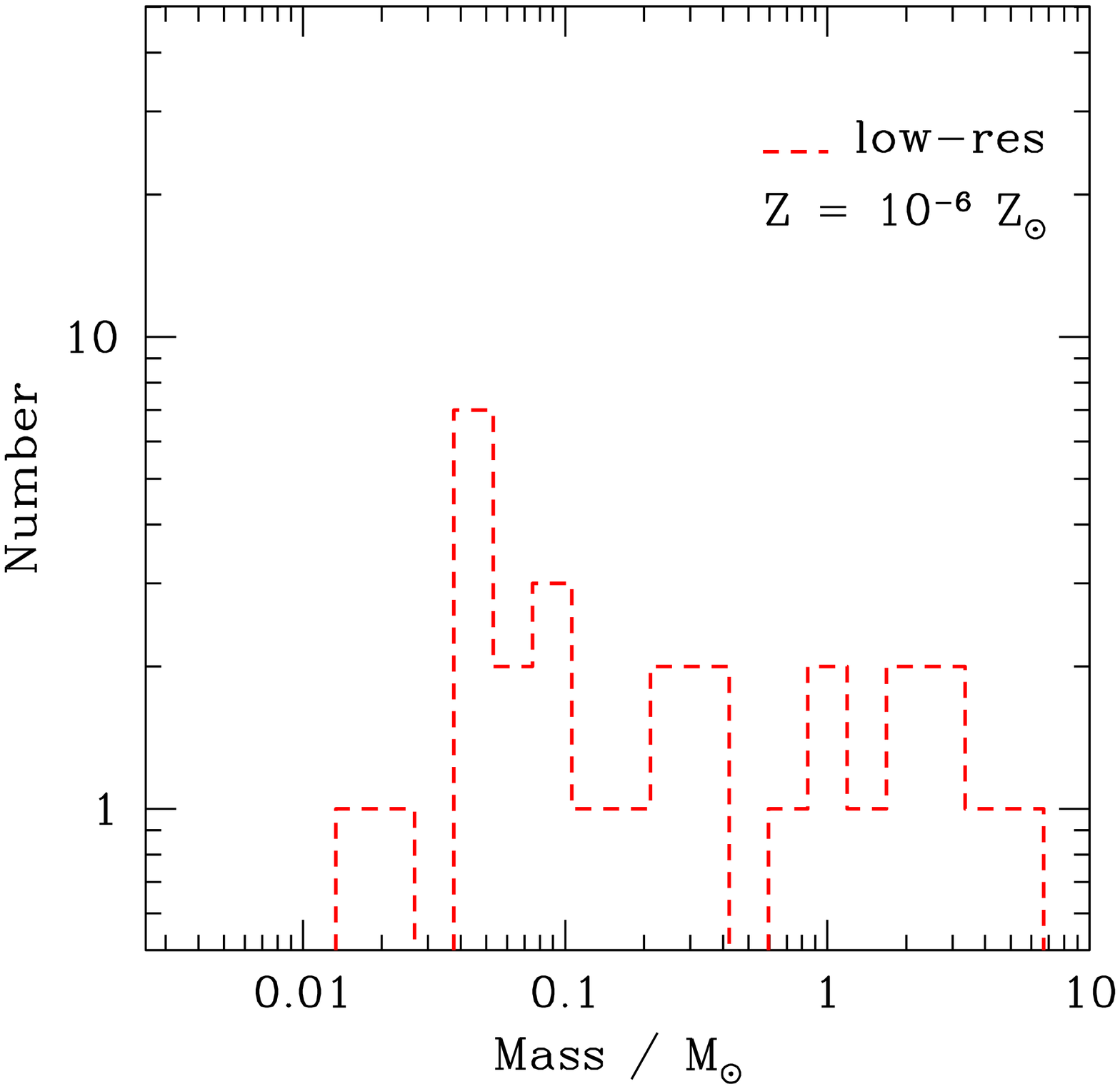}
	\includegraphics[width=1.65in,height=1.65in]{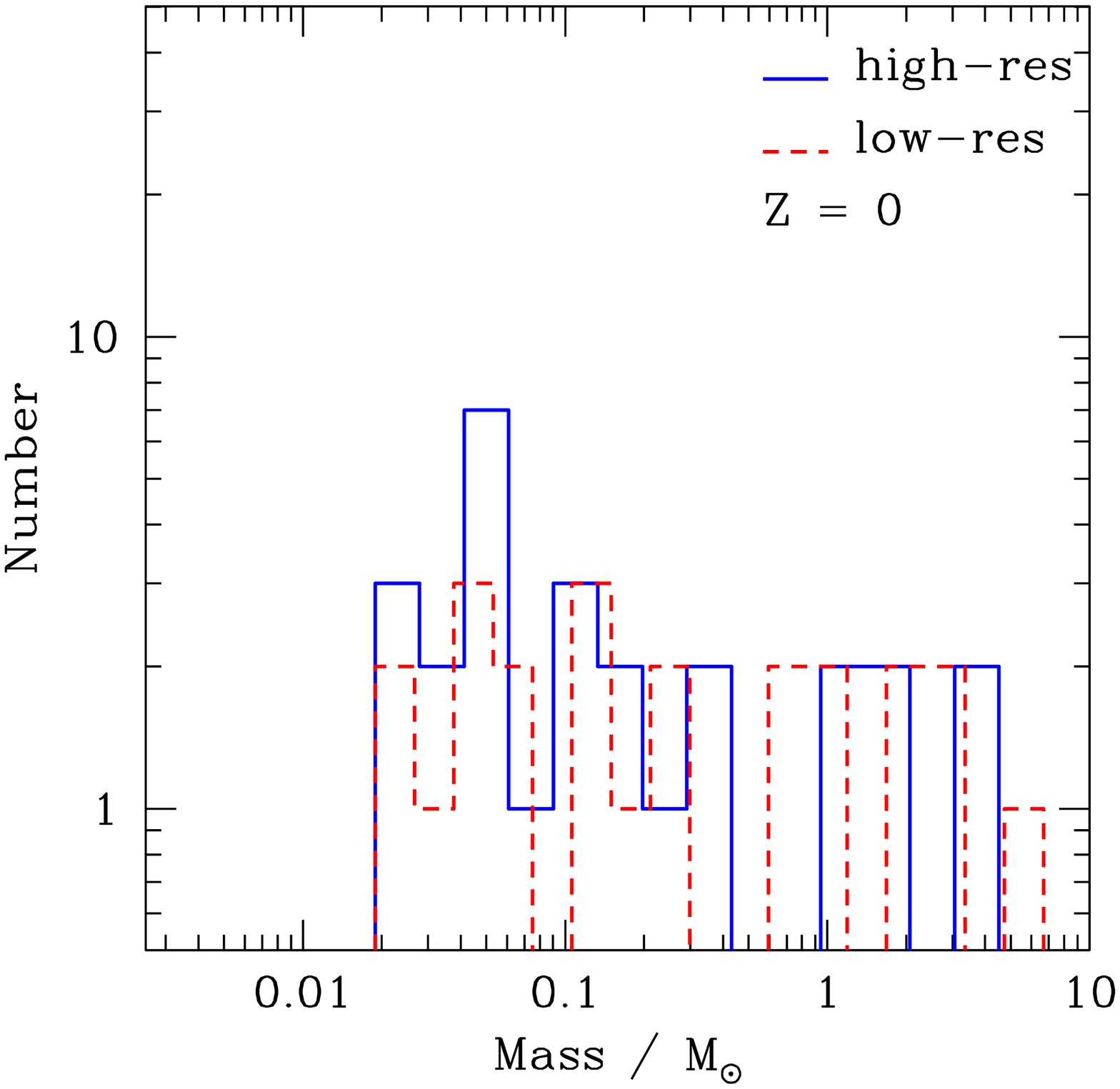}
}

\caption{\label{fig:masses}  Mass functions resulting from simulations with metallicities
${\rm Z} = 10^{-5} \: {\rm Z_{\odot}}$ (left-hand panel), ${\rm Z} = 10^{-6} \: {\rm
Z_{\odot}}$ (center panel), and ${\rm Z} = 0$ (right-hand panel). The plots refer to the
point in each simulation at which 19 \solmas of material has been accreted (which occurs at
a slightly different time in each simulation). The mass resolutions are 0.002 \solmas and
0.025 \solmas for the high and low resolution simulations, respectively.  Note the
similarity between the results of the low-resolution and high-resolution simulations. The
onset of dust-cooling in the \zfrag~cloud results in a stellar cluster which has a mass
function similar to that for present day stars, in that the majority of the mass resides in
the lower-mass objects. This contrasts with the \zsix~and primordial clouds, in which the
bulk of the cluster mass is in high-mass stars. The figure is from \cite{cgk07}.}

\end{figure}

The fragmentation of low-metallicity gas in this model is the result of two 
key features in its thermal evolution. First, the rise in the EOS curve between
densities $10^{9}$cm$^{-3}$ and $10^{11}$cm$^{-3}$ causes material to loiter at this
point in the gravitational contraction. A similar behavior at densities around $n =
10^3\,$cm$^{-3}$ is discussed by Bromm et al.\ (2001), who call it a loitering phase.
The rotationally stabilized disk-like structure, as seen in the plateau at $n \approx
10^{10}$cm$^{-3}$ in Figure \ref{fig:density}, is able to accumulate a significant
amount of mass in this phase and only slowly increases in density. Second, once the
density exceeds $n \approx 10^{12}$cm$^{-3}$, the sudden drop in the EOS curve lowers
the critical mass for gravitational collapse by two orders of magnitude. The Jeans mass
in the gas at this stage is only $M_{\rm J} = 0.01\ $M$_{\odot}$. The disk-like structure suddenly becomes highly
unstable against gravitational collapse and 
fragments vigorously on timescales of several hundred years. A very dense cluster of
embedded low-mass protostars builds up, and the protostars grow in mass by accretion
from the available gas reservoir. The number of protostars formed by the end of the 
simulation  is nearly two orders of magnitude larger than the initial number of
Jeans masses in the cloud set-up.

Because the evolutionary timescale of the system is extremely short -- the free-fall
time at a density  of $n = 10^{13}\,$cm$^{-3}$ is of the order of 10 years  --   none
of the protostars that have formed  by the time that the simulation is stopped have yet
commenced hydrogen burning. This justifies  neglecting the effects of
protostellar feedback in this study. Heating of the dust due to the significant
accretion luminosities of the newly-formed protostars will occur \citep{krum06}, but is
unlikely to be important, as the temperature of the dust at the onset of dust-induced
cooling is much higher than in a typical Galactic protostellar core ($T_{\rm dust} \sim
100 \: {\rm K}$ or more, compared to $\sim 10 \: {\rm K}$ in the Galactic case). The
rapid collapse and fragmentation of the gas also leaves no time for dynamo
amplification of magnetic fields \citep{tb04},  which in any case are expected to be
weak and dynamically unimportant in primordial and very low metallicity gas
\citep{wid02}.

The forming cluster   represents a very extreme analogue of
the clustered star formation that we know dominates in the present-day Universe
\cite[][]{ll03}. A mere 420 years after the formation of the first object, the
cluster has formed 177 stars (see Figure \ref{fig:sequence}). These occupy a region
of only around 400~AU, or $2 \times 10^{-3}$~pc, in size, roughly a hundredth of the
size of the initial cloud. With $\sim 19 \,$M$_{\odot}$  accreted at this stage, the
stellar density is $2.25 \times 10^{9}$ \solmas pc$^{-3}$. This is about five orders
of magnitude greater than the stellar density in the Trapezium cluster in Orion
\cite[][]{hh98} and about a thousand times greater than that in the core of 30
Doradus in the Large Magellanic Cloud \cite[][]{mh98}. This means that dynamical encounters 
will be extremely important during the formation of the first star cluster. The
violent environment causes stars to be thrown out of the denser regions of the cluster,
slowing down their accretion. The stellar mass spectrum thus depends on both the
details of the initial fragmentation process \cite[e.g.\ as discussed by][]{jappsen05,
clark05} as well as dynamical effects in the growing cluster \cite[][]{bcbp2001,bbv04}. 
This is different to present-day star formation, where the
situation is less clear-cut and the relative importance of these two processes may
vary strongly from region to region \cite[][]{kmk05, bb06, blz07}. 


%

%

The mass functions of the protostars at the end of the \zfrag~simulations (both high
and low resolution cases) are shown in Figure \ref{fig:masses} (left-hand panel). When
the  simulation is terminated, collapsed cores hold $\sim$ 19 \solmas of gas in
total. The mass function peaks somewhere below $0.1\ $M$_{\odot}$ and ranges from below
0.01$\,$M$_{\odot}$ to about $5\ $M$_{\odot}$. This
is not the final protostellar mass function. The continuing accretion of gas by the
cluster will alter the mass function, as will mergers between the newly-formed
protostars (which cannot be followed using our current sink particle implementation).
Protostellar feedback in the form of winds, jets and H{\sc ii} regions may also play a
role in determining the shape of the final stellar mass function. However, a key point
to note is that the chaotic evolution of a bound system such as this cluster ensures
that a wide spread of stellar masses will persist. Some stars will enjoy favourable
accretion at the expense of others that will be thrown out of the system (as can be
seen in Figure \ref{fig:sequence}), thus having their accretion effectively terminated
(see for example, the discussions in Bonnell \& Bate 2006 and Bonnell, Larson \&
Zinnecker 2007). The survival of some of the low mass  stars formed in the cluster is
therefore inevitable.
%

In  the ${\rm Z} = 0$ and \zsix~calculations \cite{cgk07} find that
fragmentation of the gas occurs as well, albeit at a much lower level than in the ${\rm Z} =
10^{-5} \: {\rm Z_{\odot}}$ run. The mass functions from these simulations are shown in
Figure \ref{fig:masses} (middle and right-hand panels), and are again taken when $\sim$
19 \solmas of gas has been accreted onto the sink particles, the same amount as is
accreted by the end of the \zfrag~calculations. Both distributions are considerably flatter than the present
day IMF, in agreement with the suggestion that Population III stars are typically very
massive. The fragmentation in the \zsix~simulation is slightly more efficient than in
the primordial case, with 33 objects forming.

\section{Conclusions}
\label{conclusions}

In this proceedings paper we have discussed several studies, where the thermodynamic behavior of the interstellar medium plays a crucial role in fragmentation and subsequent stellar birth. These examples range from star formation in the solar neighborhood at the present day to the formation of the very first stars in the early universe, and support the idea that the distribution of 
stellar masses depends, at least in part, on the thermodynamic 
state of the star-forming gas. Dips in the effective EOS, 
such that the relation between temperature $T$ and density $\rho$ changes from decreasing $T$ with increasing   $\rho$ to increasing $T$ with increasing $\rho$, i.e.\ the transition from a cooling ($\gamma < 1$) to a heating ($\gamma >1$) regime, define a characteristic mass scale for fragmentation.  

The thermodynamic state of interstellar gas is a result of the balance between 
heating and cooling processes, which in turn are determined 
by fundamental atomic and molecular physics and by chemical 
abundances. The derivation of a characteristic stellar mass can 
thus be based on quantities and constants that depend solely 
on the chemical abundances of the star forming gas. This is an attractive feature explaining the apparent universality of the IMF in the solar neighborhood as well as the transition from purely primordial high-mass star formation to the more normal low-mass mode observed today.
Clearly more work needs to be done to investigate the validity of this hypothesis.

\begin{acknowledgments}
We would like to thank Robi Banerjee, Anne-Katharina Jappsen, Richard Larson,  Yuexing Li, and  Mordecai-Mark Mac Low for stimulating discussions and collaboration.
\end{acknowledgments}


\begin{thebibliography}{}

\bibitem[Abel, Bryan, \& Norman(2002)]{abn02}
Abel, T., Bryan, G.~L., \& Norman, M.~L. 2002, Science, 295, 93

\bibitem[Beers \& Christlieb(2005)]{bc05}
Beers, T.~C., \& Christlieb, N. 2005, ARA\&A, 43, 531

\bibitem[Bonnell et al.(2001)]{bcbp2001} Bonnell, I.~A., Clarke, 
C.~J., Bate, M.~R., \& Pringle, J.~E.\ 2001, MNRAS, 324, 573 

\bibitem[Bonnell, Bate \& Vine (2004)]{bbv04} Bonnell, I.~A., Vine, 
S.~G., \& Bate, M.~R.\ 2004, MNRAS, 349, 735 

\bibitem[Bonnell \& Bate(2006)]{bb06}
Bonnell, I.~A., \& Bate, M.~R. 2006, MNRAS, 370, 488

\bibitem[Bonnell, Larson \& Zinnecker (2007)]{blz07} Bonnell, I.~A., Larson, 
R.~B., \& Zinnecker, H.\ 2007, Protostars and Planets V,
B. Reipurth, D. Jewitt, and K. Keil (eds.), University of Arizona Press, Tucson, p.149 

\bibitem[Bromm et~al.(2001)]{bfcl01}
Bromm, V., Ferrara, A., Coppi, P.~S., \& Larson, R.~B. 2001, MNRAS, 328, 969

\bibitem[Bromm, Coppi, \& Larson(2002)]{bcl02}
Bromm, V., Coppi, P.~S., \& Larson, R.~B. 2002, ApJ, 564, 23

\bibitem[Bromm \& Loeb(2003)]{bl03}
Bromm, V. \& Loeb, A. 2003, Nature, 425, 812

\bibitem[Bromm \& Loeb(2004)]{bl04}
Bromm, V., \& Loeb, A. 2004, New Astron., 9, 353

\bibitem[Chabrier(2003)]{chabrier03}
Chabrier, G. 2003, PASP, 115, 763 

\bibitem[Christlieb et~al.(2002)]{christlieb02} 
Christlieb, N., Bessell, M. S., Beers, T. C., Gustafsson, B., Korn, A., Barklem, P. S., Karlsson, T., 
Mizuno-Wiedner, M., \& Rossi, S.  2002, Nature, 419, 904 

\bibitem[Clark \& Bonnell(2005)]{clark05} Clark, P.~C., \& 
Bonnell, I.~A.\ 2005, MNRAS, 361, 2 

\bibitem[Clark, Glover, \& Klessen(2007)]{cgk07}
Clark, P.~C., Glover, S.~C.~O., \& Klessen, R.~S. 2007, ApJ, in press; arXiv:0706.0613

\bibitem[{{Evans}(1999)}]{EVA99}
{Evans}, N.~J. 1999, ARAA, 37, 311

\bibitem[{{Evans} {et~al.}(2001){Evans}, {Rawlings}, {Shirley}, \&
  {Mundy}}]{EVA01}
{Evans}, N.~J., {Rawlings}, J.~M.~C., {Shirley}, Y.~L., \& {Mundy}, L.~G. 2001,
  ApJ, 557, 193

\bibitem[Frebel, Johnson, \& Bromm(2007)]{fjb07}
Frebel, A., Johnson, J.~L., \& Bromm, V. 2007, astro-ph/0701395

\bibitem[Glover(2005)]{glover05}
Glover, S.~C.~O. 2005, Space \ Sci.\ Reviews, 117, 445

\bibitem[Glover \& Mac Low(2007)]{GM07}
Glover, S.~C.~O., Mac Low, M.-M. 2007, ApJ, 659, 1317

\bibitem[{{Hayashi}(1966)}]{HAY66}
{Hayashi}, C. 1966, ARAA, 4, 171

\bibitem[{{Hayashi} \& {Nakano}(1965)}]{HAY65}
{Hayashi}, C. \& {Nakano}, T. 1965, Prog.~Theor.~Phys., 34, 754

\bibitem[Heger \& Woosley(2002)]{hw02}
Heger, A., Woosley, S.~E. 2002, ApJ, 567, 532

\bibitem[Hillenbrand \& Hartmann(1998)]{hh98} 
Hillenbrand, L.~A., \& Hartmann, L.~W.\ 1998, ApJ, 492, 540 

\bibitem[Jappsen et~al.(2005)]{jappsen05} 
Jappsen, A.-K., Klessen, R.~S., Larson, R.~B., Li, Y., \& Mac Low, M.-M.\ 2005, A\&A, 435, 
611 

\bibitem[Komiya et~al.(2007)]{kom07}
Komiya, Y., Suda, T., Minaguchi, H., Shigeyama, T., 
Aoki, W., \& Fujimoto, M.~Y. 2007, ApJ, 658, 367

\bibitem[{{Koyama} \& {Inutsuka}(2000)}]{KOY00}
{Koyama}, H. \& {Inutsuka}, S. 2000, ApJ, 532, 980


\bibitem[Kroupa(1998)]{kroupa98} Kroupa, P.\ 1998, MNRAS, 298, 
231

\bibitem[Kroupa(2002)]{kroupa02} Kroupa, P.\ 2002, Science, 295, 
82 

\bibitem[Krumholz, McKee, \& Klein(2005)]{kmk05}
Krumholz, M.~R., McKee, C.~F., \& Klein, R.~I. 2005, Nature, 438, 332

\bibitem[Krumholz(2006)]{krum06}
Krumholz, M.~R. 2006, ApJ, 641, L45

\bibitem[Lada \& Lada(2003)]{ll03} 
Lada, C.~J., \& Lada, E.~A.\ 2003, ARA\&A, 41, 57 

\bibitem[{{Larson}(1969)}]{LAR69}
{Larson}, R.~B. 1969, MNRAS, 145, 271

\bibitem[{{Larson}(1973{\natexlab{b}})}]{LAR73}
---. 1973{\natexlab{b}}, Fundamentals of Cosmic Physics, 1, 1

\bibitem[{{Larson}(1985)}]{LAR85}
---. 1985, MNRAS, 214, 379

\bibitem[{Larson}(2005)]{LAR05}
---. 2005, MNRAS, 359, 211

\bibitem[{{Li} {et~al.}(2003){Li}, {Klessen}, \& {Mac Low}}]{LI03}
{Li}, Y., {Klessen}, R.~S., \& {Mac Low}, M.-M. 2003, ApJ, 592, 975

\bibitem[Loeb \& Barkana(2001)]{lb01}
Loeb, A., Barkana, R. 2001, ARA\&A, 39, 19

\bibitem[{{Low} \& {Lynden-Bell}(1976)}]{LOW76}
{Low}, C. \& {Lynden-Bell}, D. 1976, MNRAS, 176, 367

\bibitem[Lucatello et~al.(2005)]{luc05}
Lucatello, S., Tsangarides, S., Beers, T.~C., Carretta, E.,
Gratton, R.~G., \& Ryan, S.~G. 2005, ApJ, 625, 825

\bibitem[Massey \& Hunter(1998)]{mh98} 
Massey, P., \& Hunter, D.~A.\ 1998, ApJ, 493, 180

\bibitem[{{Masunaga} \& {Inutsuka}(2000)}]{MAS00}
{Masunaga}, H. \& {Inutsuka}, S. 2000, ApJ, 531, 350

  
  \bibitem[{{Myers}(1978)}]{MYE78}
{Myers}, P.~C. 1978, ApJ, 225, 380

\bibitem[Omukai et~al.(2005)]{om05}
Omukai, K., Tsuribe, T., Schneider, R., \& Ferrara, A. 2005, ApJ, 626, 627

\bibitem[O'Shea \& Norman(2007)]{oshn07}
O'Shea, B.~W., \& Norman, M.~L. 2007, ApJ, 654, 66

\bibitem[Ryan et~al.(2005)]{ry05}
Ryan, S.~G., Aoki, W., Norris, J.~E., \& Beers, T.~C. 2005, ApJ, 635, 349

\bibitem[Santoro \& Shull(2006)]{san06}
Santoro, F. \& Shull, J.~M. 2006, ApJ, 643, 26

\bibitem[{{Scalo}(1998)}]{scalo98}
{Scalo}, J. 1998, in ASP Conf. Ser. 142: The Stellar Initial Mass Function
  (38th Herstmonceux Conference), ed. G. Gilmore \& D. Howell (San Francisco:
  Astron. Soc. Pac.), 201
  
\bibitem[Schneider et~al.(2002)]{sch02}
Schneider, R., Ferrara, A., Natarajan, P. \& Omukai, K. 2002, ApJ, 571, 30

\bibitem[Schneider et~al.(2006)]{sch06}
Schneider, R., Omukai, K., Inoue, A.~K., \& Ferrara, A. 2006, MNRAS, 369, 1437

\bibitem[{{Tafalla} {et~al.}(2004){Tafalla}, {Myers}, {Caselli}, \&
  {Walmsley}}]{TAF04}
{Tafalla}, M., {Myers}, P.~C., {Caselli}, P., \& {Walmsley}, C.~M. 2004, A{\&}A,
  416, 191

\bibitem[Tan \& Blackman(2004)]{tb04}
Tan, J.~C., \& Blackman, E.~G. 2004, ApJ, 603, 401

\bibitem[Tsuribe \& Omukai(2006)]{to06}
Tsuribe, T., \& Omukai, K. 2006, ApJ, 642, L61

\bibitem[Widrow(2002)]{wid02}
Widrow, L.~M. 2002, Rev.\ Mod.\ Phys., 74, 775

\bibitem[Yoshida {\em et~al.}(2006)]{yoha06}
Yoshida, N., Omukai, K., Hernquist, L., \& Abel, T. 2006, ApJ, 652, 6

\bibitem[{{Zucconi} {et~al.}(2001){Zucconi}, {Walmsley}, \& {Galli}}]{ZUC01}
{Zucconi}, A., {Walmsley}, C.~M., \& {Galli}, D. 2001, A{\&}A, 376, 650




\end{thebibliography}
\end{document}